\documentclass{article}

\PassOptionsToPackage{numbers}{natbib}
\usepackage[final]{neurips_ts4h_2022}

\usepackage[utf8]{inputenc} 
\usepackage[T1]{fontenc}    
\usepackage{hyperref}       
\usepackage{url}            
\usepackage{booktabs}       
\usepackage{amsfonts}       
\usepackage{nicefrac}       
\usepackage{microtype}      
\usepackage{xcolor}         
\usepackage{graphicx, subfig}
\usepackage{wrapfig}

\title{Automatic Sleep Scoring from Large-scale Multi-channel Pediatric EEG}

\author{%
  Harlin Lee \\
  Department of Mathematics\\
  University of California Los Angeles\\
  Los Angeles, CA 90095\\
  \texttt{harlin@math.ucla.edu} \\
   \And
   Aaqib Saeed\\
   Philips Research\\
   Eindhoven, Netherlands\\
   \texttt{aaqib.saeed@philips.com} \\
}

\begin{document}

\maketitle

\begin{abstract}
 Sleep is particularly important to the health of infants, children, and adolescents, and sleep scoring is the first step to accurate diagnosis and treatment of potentially life-threatening conditions. But pediatric sleep is severely under-researched compared to adult sleep in the context of machine learning for health, and sleep scoring algorithms developed for adults usually perform poorly on infants. Here, we present the first automated sleep scoring results on a recent large-scale pediatric sleep study dataset that was collected during standard clinical care. We develop a transformer-based model that learns to classify five sleep stages from millions of multi-channel electroencephalogram (EEG) sleep epochs with 78\% overall accuracy. Further, we conduct an in-depth analysis of the model performance based on patient demographics and EEG channels. The results point to the growing need for machine learning research on pediatric sleep.
\end{abstract}

\section{Introduction}

Sleep is necessary for everyone, but it is particularly important to the health and development of infants, children and adolescents. Sleep disorders or sleep disturbances can negatively affect one's cognitive and physical functions, and even lead to serious medical conditions. 
For example, obstructive sleep apnea (OSA) contributes to neurobehavioral issues \citep{beebe_neuropsychological_2004,zhao2018association} and morbidity \citep{jennum2013morbidity,lumeng_epidemiology_2008} in infants and children. Yet, pediatric sleep is severely under-researched compared to adult sleep in the context of machine learning for health and well-being, due to the following:
\begin{itemize}
    \item Sleep disturbances in children tend to be under-reported \citep{blunden2004sleep,grandner2017insomnia,schreck2011knowledge}, downplaying the clinical need for novel approaches in this field.
    \item Many (including clinicians) assume that children are ``just little adults'' when in fact pediatric sleep is physiologically distinct from adult sleep \citep{accardo2010differences,owens2012pro,sun2017revisiting}, and therefore computational models based on adult sleep do not generalize well to pediatric sleep.
    \item Benchmark datasets are the backbone of machine learning research \citep{paullada2021data,49953,vanschoren2021announcing}, but a large, high-quality dataset dedicated to pediatric sleep has been published only recently \citep{ibrahim2021best,lee2022large}.
\end{itemize}
This paper brings attention to the growing need for machine learning research on pediatric sleep by focusing on \textit{automated pediatric sleep scoring}, which has been overlooked in favor of automated adult sleep scoring by the community. Diagnoses of many sleep conditions require polysomnography (PSG), or overnight sleep study, where a patient sleeps in a clinic while their physiological signals are monitored under the supervision of trained technicians \citep{berry_aasm_2017,berry_aasm_2018,kushida_practice_2005}. A PSG dataset may include modalities such as electroencephalogram (EEG), electromyelogram (EMG), electrooculogram (EOG), and respiratory airflow. 
A crucial first step towards diagnosis with PSG is sleep scoring, or sleep stage classification, which assigns every 30-second segment of sleep into two stages, rapid eye movement (REM), and non-REM, then further divides the latter into shallow sleep (stages N1 and N2) and deep sleep (stage N3). In a typical clinical setting, this process is done manually by a technician, which is highly labor-intensive, time-consuming and prohibitively expensive. 

Naturally there have been many attempts to automate sleep scoring, especially in recent years with the help of deep neural networks and freely-available public PSG datasets; see reviews in \citet{bandyopadhyay2022clinical,fiorillo2019automated, phan2021automatic,watson2021artificial}. 
In particular, several works \citep{chen2021end, kim2021sleep,phan2022sleeptransformer, yang2021exploring} have attempted to utilize transformer-based models for processing EEG signals and achieve superior performance over other classic deep architectures. Given that, we design a simple yet effective neural architecture that can process millions of multi-channel EEG signals and learn useful representations for pediatric sleep scoring. Our model is based on the transformer architecture that operates directly on patches as input and maintains the same resolution and representations throughout all layers. However, our model differs from previous transformer-based approaches to sleep-stage scoring in that: 1) it is trained specifically for pediatric sleep scoring; 2) it does not utilize any other modalities except EEG signals as other works employ additional modalities, e.g., EOG; and 3) it directly operates over raw signals as opposed to time-frequency images to further simplify the learning pipeline and improving training efficiency.

We develop and demonstrate our model on the new Nationwide Children's Hospital (NCH) Sleep DataBank \citep{lee2022large}, which has not been explored in sleep scoring literature before. This massive dataset allows us to leverage the full power of deep learning. It explicitly focuses on pediatric sleep, and the sleep studies were conducted in a current real-world clinical setting (i.e. in-the-wild in NCH between 2017 and 2019). Hence our model is trained from EEGs that are closest to what it will see in future deployment, which is unlike prior work on sleep scoring that learn from mostly healthy adults in a clinical trial. Our transformer-based model achieves an overall pediatric sleep scoring accuracy of 78.2\%, and our analysis reveals that the accuracy is above 80\% for 6-15 year old patients. We believe that the difference in performance between age groups supports our call for dedicated attention to pediatric sleep from the machine learning for health community.

\section{Approach}
We develop a neural network model for predicting sleep stages in a real-world clinical environment from pediatric multi-channel EEG signals. We design a patch-based transformer model that operates over one-second segments of sleep, which provides strong support for long-range modeling dependencies in the input signal to learn discriminative features.

Our transformer-based model is inspired by the ViT~\citep{dosovitskiy2020image} network, which we adapt here to multi-channel time-series signals. The model accepts inputs of the shape \texttt{(Sampling frequency in Hz $\times$ \# of seconds) $\times$  (\# of EEG channels)} $= 3,840 \times 7$, after which the instance normalization layer normalizes each EEG signal channel-wise independently. The patch generation layer then splits the sleep epoch by every second, creating $30$ patches of input with shape $128 \times 7$. This is analogous to tokenization in natural language processing (NLP), where a piece of text is converted into smaller units (i.e. tokens) such as words or characters. This helps the transformer learn which seconds of the EEG signals are important for sleep scoring.

After the patches are generated, they are embedded into $64$-dimensional vectors via a linear patch encoder layer. This is then added to $64$-dimensional positional vectors to create images that encode both positional and waveform shape information of the input patches. The rest of the model is similar to a classic transformer encoder with $8$ blocks with $4$ attention heads, which is explained in more detail in Section \ref{subsec_appendix:transformer}. Each block has a normalization layer, a multi-head attention layer, another normalization layer, and a two-layer multi-layer perceptron (MLP) with $128$ and $64$ units. For feature aggregation, we use global average pooling followed by a classification layer with units equal to the number of sleep stages, i.e. $5$.

\section{Results}
We utilize the NCH SleepBank dataset, which comprises approximately $3.6$ million fully-annotated EEG examples by domain experts, for training and evaluating models. Only seven-channel EEG signals (F4-M1, O2-M1, C4-M1, O1-M2, F3-M2, C3-M2, and CZ-01) at $128$ Hz are used to classify instances into five sleep stages (i.e., wakefulness, non-REM stages 1, 2, 3, REM). Detailed information about the NCH dataset, including patient characteristics and annotation strategy, is provided in supplementary materials Section~\ref{subsec_appendix:data}. To evaluate model performance, we compute precision, recall, F1-score, and accuracy based on the confusion matrix, and also assess generalization across age groups, races, and gender. Finally, we perform ablation over EEG channels to estimate the contribution of each channel toward sleep scoring.   

\subsection{Data preparation}
We use 3,928 PSGs from $3,631$ unique patients for model training and evaluation. In particular, we split the patients into $70\%, 10\%$, and $20\%$ for training, validation, and testing, respectively, so that the three splits have no overlap in patients. During the learning phase, we monitor the validation set performance for model checkpointing, and report results on the test set. Our training set consists of $2.5$+ million instances, and the test set has $730$K+ instances, as shown in Table \ref{tab:dataset_size} of Section~\ref{subsec_appendix:data}. To the best of our knowledge, we, for the first time, report results on a large-scale pediatric sleep stage scoring dataset that is collected in the wild. We provide the rest of the data pre-processing and related information in Section \ref{subsec_appendix:data}, including patient demographic characteristics in Table \ref{tbl:patient_demographics}.

\subsection{Model demonstrates strong pediatric sleep scoring performance}

\begin{figure*}[!htp]
\centering
\subfloat[Confusion Matrix]{\includegraphics[width=0.5\textwidth]{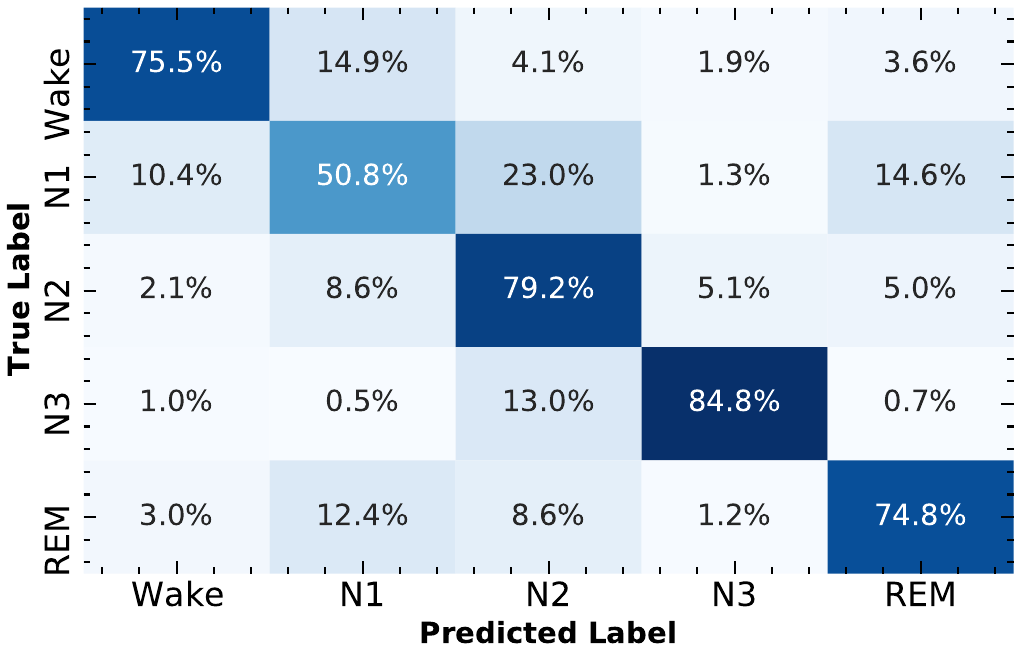}
\label{fig:confusion_mat_t}
}
\subfloat[Performance Metrics]{\includegraphics[width=0.51\textwidth]{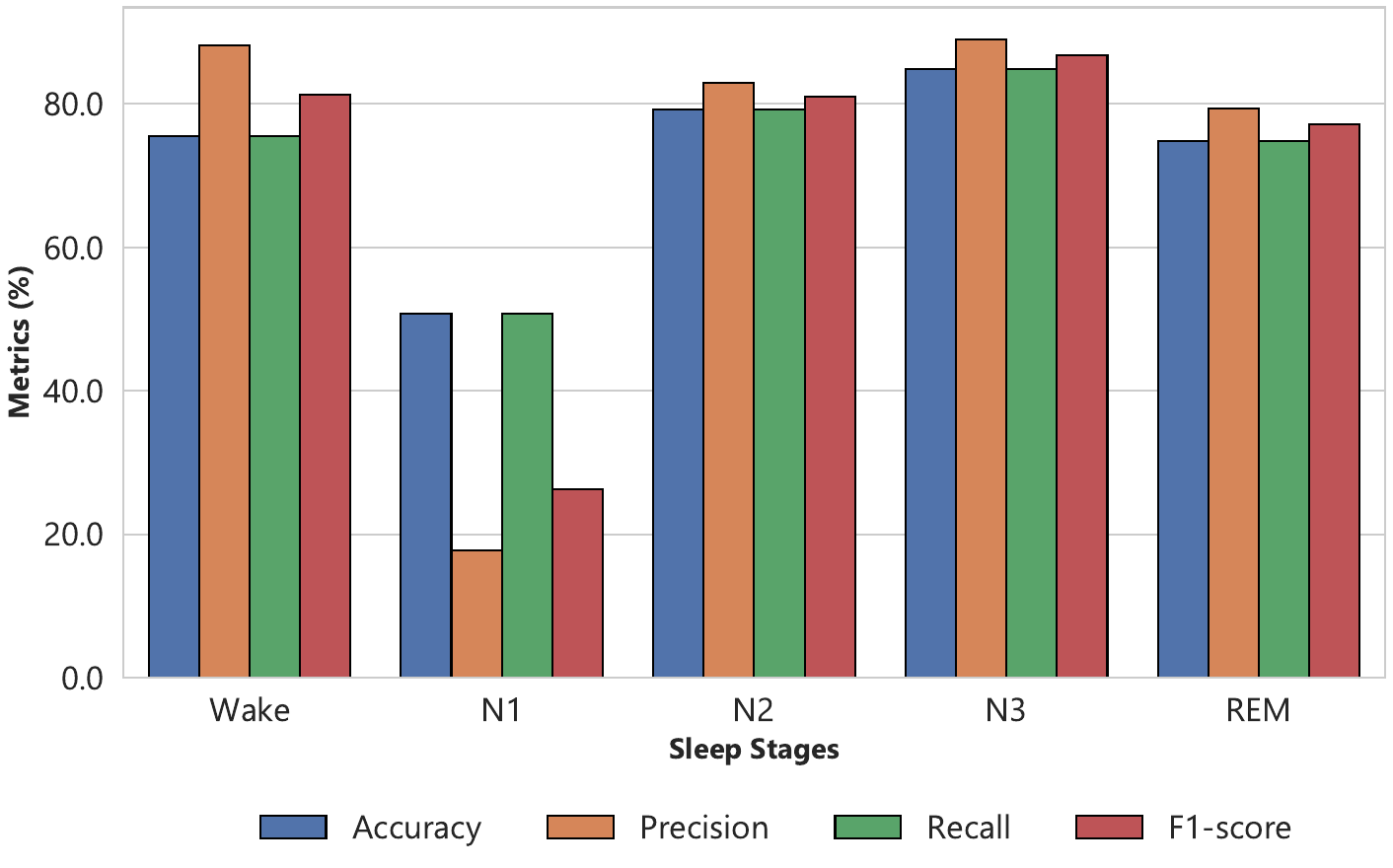}
\label{fig:classification_metrics_t}
} 
\caption{a) Normalized confusion matrix for sleep scoring on the entire test set. The number in $i$th row and $j$th column indicates the percentage (\%) of samples in stage $i$ (according to manual scoring) that were predicted to be in stage $j$ by our classifier. Each row adds to $100$\%. Overall accuracy of our model across all sleep stages is $78.2$\%. b) Model performances on the entire test set, as evaluated by accuracy, precision, recall, and F1-score (weighted).}
\end{figure*}

Across all sleep stages in the test set, the transformer model achieves $78.2$\% accuracy, F1-score (macro) of $70.5$\%, F1-score (weighted) of $79.9$\%, and Cohen's Kappa score of $71.0$\%. Model performance for each sleep stage is presented in Figure \ref{fig:confusion_mat_t} as a normalized confusion matrix, and in Figure \ref{fig:classification_metrics_t} in terms of accuracy, precision, recall, and F1-score. The model demonstrates strong predictive power (near $80$\%) for Wake, N2, N3, and REM, but not as much in predicting N1, which has the smallest sample size. The model has lower precision for sleep epochs in N1, often inaccurately labeling them as N2 or REM. Nonetheless, this is a huge improvement over the wavelet-based baseline classifier in \citet{lee2022large}, which had $64.4$\% accuracy across all sleep stages and only $0.9$\% with N1. 

We also visualize the features learned by the transformer model in Figure \ref{fig:tsne_features} in Section~\ref{subsec_appendix:tsne}, projecting them from 128-dimensional to 2-dimensional space via t-SNE \citep{van2008visualizing}. The clusters that naturally form for each sleep stage suggest that the transformer model learns meaningful features from the raw EEG signals before entering the final classifier layer. Furthermore, we note that N3 samples seem to be most well-separated, while N1 samples seem to overlap with other stages the most, which aligns with the classification accuracy results in Figure \ref{fig:confusion_mat_t}.

\subsection{Model sleep scores better on 6 to 15 year olds and children of Asian, Others and Unknown race with over $80$\% accuracy}

\begin{table}[h]
\centering
\setlength\tabcolsep{4pt}
\caption{Transformer model performance on different racial and groups and sex in the test set. Others and Unknown race is defined identically to Table \ref{tbl:patient_demographics}. F1 refers to weighted F1-score.}
\begin{footnotesize}
\begin{tabular}{lrr} 
\toprule
&  Accuracy (\%) & F1-score (\%)  \\
\midrule \addlinespace[1mm]
 \textbf{Race} && \\\addlinespace[1mm]
~White &  78.6 &  80.3 \\\addlinespace[1mm]
~Black or African American & 76.4 & 78.3 \\\addlinespace[1mm]
~Multiple Races & 78.0 & 79.6 \\\addlinespace[1mm]
~Asian & 78.7 & 80.3 \\\addlinespace[1mm]
~Others and Unknown & 80.6 & 82.6 \\\addlinespace[2mm]
\textbf{Sex} & & \\ \addlinespace[1mm]
~Male & 77.9 & 79.6\\\addlinespace[1mm]
~Female &78.5 & 80.4\\ 
 \bottomrule
\end{tabular}
\end{footnotesize}
\label{tbl:transformer_race} 
\end{table}

\begin{figure}[h]
\centering
    \includegraphics[width=0.5\linewidth]{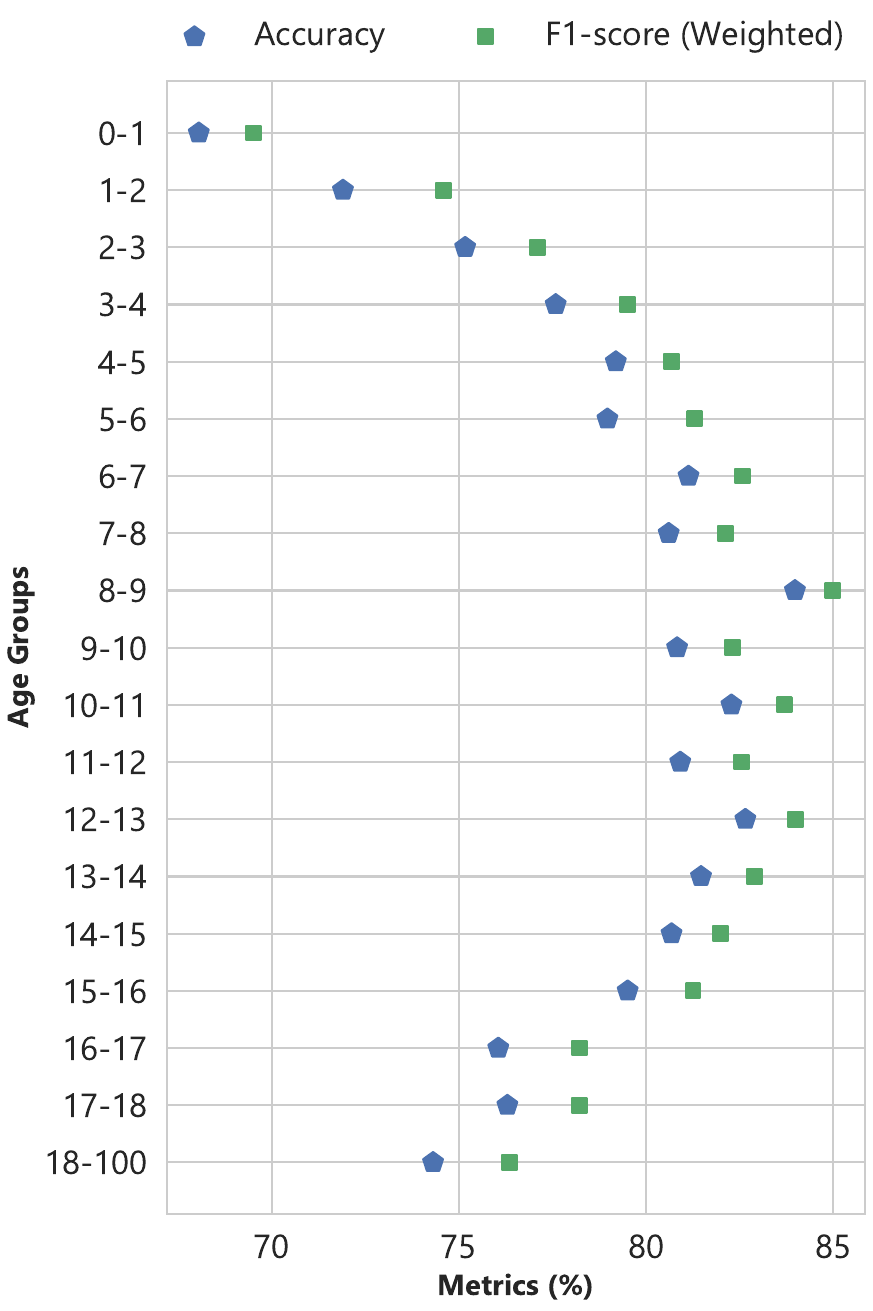}
    \caption{Performance comparison of transformer model on different age groups in the test set, as measured by accuracy and weighted F1-score.}
    \label{fig:transformer_age}
\end{figure}

Table \ref{tbl:transformer_race} and Figure \ref{fig:transformer_age} report the transformer model performance on different subsets of the patients. Figure \ref{fig:transformer_age} shows that the model achieves the highest accuracy ($85$\%) on the 8 to 9 year olds, and the lowest ($70$\%) on infants less than 1 year old. From 6 to 15 year old age groups, the classification accuracy is above $80$\%, and subsequently higher than the model's average accuracy across age groups. In terms of race, the model achieves the highest accuracy (about $81$\%) on Others and Unknown, and lowest accuracy (about $76$\%) on Black or African Americans. Finally, we observed slightly better performance on female patients.

\subsection{Predictive power is not from a single EEG channel}
Next, we perform an experiment to determine the individual contributions of the EEG channels towards sleep scoring. Seven identical transformer models are created according to the description in Section~\ref{subsec_appendix:transformer}. Then, each model is trained using only one of the seven EEG channels. For example, the first row of Table \ref{tbl:ablation_channels} shows the classification accuracy of a transformer model that only had access to the F4-M1 channel EEG during both training and testing. None of the seven models is able to achieve the results of the original transformer model, lending support to the use of multi-channel EEG signals. However, the model trained on F3-M2 channel achieves highest accuracy in classifying sleep stages Wake, N1, and N3, while the C3-M2 channel model does so for N2. Finally, the F4-M1 channel model demonstrates a markedly improved performance in identifying REM stages.

\begin{table}[thp]
\centering
\caption{Classification accuracy (\%) on test set for transformer models trained on single EEG channels. The highest accuracy for each sleep stage (column) is bolded.} 
\begin{footnotesize}
\begin{tabular}{lrrrrrr} 
\toprule
\textbf{Channel} &\multicolumn{6}{c}{\textbf{Sleep Stage}}   \\ \addlinespace[1mm]
 & Wake & N1 & N2 & N3 & REM & All \\
\midrule \addlinespace[1mm]
F4-M1 & 69.1 & 39.8 & 75.0 & 83.1 & \textbf{78.0} & 75.1 \\\addlinespace[1mm]
O2-M1 & 70.1 & 31.5 & 74.1 & 81.4 & 60.7 & 71.4 \\\addlinespace[1mm]
C4-M1 & 68.2 & 39.9 & 77.0 & 83.5 & 70.8 & 74.6 \\\addlinespace[1mm]
O1-M2 & 67.7 & 35.0 & 73.8 & 78.7 & 67.4 & 71.4 \\\addlinespace[1mm]
F3-M2 & \textbf{72.4} & \textbf{41.8} & 75.5 &  \textbf{84.2} & 71.1 & 75.1 \\\addlinespace[1mm]
C3-M2 & 72.2 & 34.7 & \textbf{78.3} & 83.8 & 70.8 & \textbf{75.7} \\\addlinespace[1mm]
CZ-O1 & 69.1 & 34.6 & 76.5 & 80.1 & 66.5 & 72.9 \\
 \bottomrule
\end{tabular}
\end{footnotesize}
\label{tbl:ablation_channels} 
\end{table}

\section{Conclusions}
We developed and trained a transformer model on more than 3,900 recent pediatric sleep studies collected during standard hospital care. The model predicted $5$ sleep stages (Wake, N1, N2, N3, REM) from $7$ raw EEG channels with $78.2$\% accuracy, which is the highest accuracy reported for automatic sleep scoring on such a large-scale pediatric dataset to the best of our knowledge.

We believe this work sheds light on the importance and challenges of machine learning for pediatric sleep, as well as many future research ideas. First, the challenge in predicting the infrequent N1 stages, while consistent with previous literature, remains an open problem. Prediction performance for infants less than one year old also has room for improvement; in fact, developing models separately for neonates, children, and adolescents may be of interest. Finally, as the NCH Sleep DataBank also provides the patients' electronic health records, we plan to build on this work to develop diagnostic models for sleep disorders.

\begin{ack}
The work of Harlin Lee was partially supported by NSF DMS-195233.
\end{ack}

\bibliographystyle{abbrvnat}
\bibliography{main}

\begin{thebibliography}{31}
\providecommand{\natexlab}[1]{#1}
\providecommand{\url}[1]{\texttt{#1}}
\expandafter\ifx\csname urlstyle\endcsname\relax
  \providecommand{\doi}[1]{doi: #1}\else
  \providecommand{\doi}{doi: \begingroup \urlstyle{rm}\Url}\fi

\bibitem[Accardo et~al.(2010)Accardo, Shults, Leonard, Traylor, and
  Marcus]{accardo2010differences}
J.~A. Accardo, J.~Shults, M.~B. Leonard, J.~Traylor, and C.~L. Marcus.
\newblock Differences in overnight polysomnography scores using the adult and
  pediatric criteria for respiratory events in adolescents.
\newblock \emph{Sleep}, 33\penalty0 (10):\penalty0 1333--1339, 2010.

\bibitem[Bandyopadhyay and Goldstein(2022)]{bandyopadhyay2022clinical}
A.~Bandyopadhyay and C.~Goldstein.
\newblock Clinical applications of artificial intelligence in sleep medicine: a
  sleep clinician’s perspective.
\newblock \emph{Sleep and Breathing}, pages 1--17, 2022.

\bibitem[Beebe et~al.(2004)Beebe, Wells, Jeffries, Chini, Kalra, and
  Amin]{beebe_neuropsychological_2004}
D.~W. Beebe, C.~T. Wells, J.~Jeffries, B.~Chini, M.~Kalra, and R.~Amin.
\newblock Neuropsychological effects of pediatric obstructive sleep apnea.
\newblock \emph{J. Int. Neuropsychol. Soc.}, 10\penalty0 (7):\penalty0 962,
  2004.

\bibitem[Berry et~al.(2017)Berry, Brooks, Gamaldo, Harding, Lloyd, Quan,
  Troester, and Vaughn]{berry_aasm_2017}
R.~B. Berry, R.~Brooks, C.~Gamaldo, S.~M. Harding, R.~M. Lloyd, S.~F. Quan,
  M.~T. Troester, and B.~V. Vaughn.
\newblock \emph{The {AASM} {Manual} for the scoring of {Sleep} and {Associated}
  {Events}: {Rules}, {Terminology} and {Technical} {Specifications}. {Version}
  2.4.}
\newblock American Academy of Sleep Medicine, Darien, IL, 2017.

\bibitem[Berry et~al.(2018)Berry, Albertario, Harding, Lloyd, Plante, Quan,
  Troester, and Vaughn]{berry_aasm_2018}
R.~B. Berry, C.~L. Albertario, S.~M. Harding, R.~M. Lloyd, D.~T. Plante, S.~F.
  Quan, M.~M. Troester, and B.~V. Vaughn.
\newblock \emph{The {AASM} {Manual} for the scoring of {Sleep} and {Associated}
  {Events}: {Rules}, {Terminology} and {Technical} {Specifications}. {Version}
  2.5.}
\newblock American Academy of Sleep Medicine, Darien, IL, 2018.

\bibitem[Blunden et~al.(2004)Blunden, Lushington, Lorenzen, Ooi, Fung, and
  Kennedy]{blunden2004sleep}
S.~Blunden, K.~Lushington, B.~Lorenzen, T.~Ooi, F.~Fung, and D.~Kennedy.
\newblock Are sleep problems under-recognised in general practice?
\newblock \emph{Archives of disease in childhood}, 89\penalty0 (8):\penalty0
  708--712, 2004.

\bibitem[Chen et~al.(2021)Chen, Yang, Wang, Huang, Ono, Altaf-Ul-Amin, and
  Kanaya]{chen2021end}
Z.~Chen, Z.~Yang, D.~Wang, M.~Huang, N.~Ono, M.~Altaf-Ul-Amin, and S.~Kanaya.
\newblock An end-to-end sleep staging simulator based on mixed deep neural
  networks.
\newblock In \emph{2021 IEEE International Conference on Bioinformatics and
  Biomedicine (BIBM)}, pages 848--853. IEEE, 2021.

\bibitem[Dosovitskiy et~al.(2020)Dosovitskiy, Beyer, Kolesnikov, Weissenborn,
  Zhai, Unterthiner, Dehghani, Minderer, Heigold, Gelly,
  et~al.]{dosovitskiy2020image}
A.~Dosovitskiy, L.~Beyer, A.~Kolesnikov, D.~Weissenborn, X.~Zhai,
  T.~Unterthiner, M.~Dehghani, M.~Minderer, G.~Heigold, S.~Gelly, et~al.
\newblock An image is worth 16x16 words: Transformers for image recognition at
  scale.
\newblock \emph{arXiv preprint arXiv:2010.11929}, 2020.

\bibitem[Fiorillo et~al.(2019)Fiorillo, Puiatti, Papandrea, Ratti, Favaro,
  Roth, Bargiotas, Bassetti, and Faraci]{fiorillo2019automated}
L.~Fiorillo, A.~Puiatti, M.~Papandrea, P.-L. Ratti, P.~Favaro, C.~Roth,
  P.~Bargiotas, C.~L. Bassetti, and F.~D. Faraci.
\newblock Automated sleep scoring: A review of the latest approaches.
\newblock \emph{Sleep medicine reviews}, 48:\penalty0 101204, 2019.

\bibitem[Grandner and Chakravorty(2017)]{grandner2017insomnia}
M.~A. Grandner and S.~Chakravorty.
\newblock Insomnia in primary care: Misreported, mishandled, and just plain
  missed.
\newblock \emph{Journal of Clinical Sleep Medicine}, 13\penalty0 (8):\penalty0
  937--939, 2017.

\bibitem[Ibrahim et~al.(2021)Ibrahim, Stone, and Rosen]{ibrahim2021best}
S.~Ibrahim, J.~Stone, and C.~L. Rosen.
\newblock Best practices for accommodating children in the polysomnography lab:
  Enhancing quality and patient experience.
\newblock In \emph{Pediatric Sleep Medicine}, pages 169--177. Springer, 2021.

\bibitem[Jennum et~al.(2013)Jennum, Ibsen, and Kjellberg]{jennum2013morbidity}
P.~Jennum, R.~Ibsen, and J.~Kjellberg.
\newblock Morbidity and mortality in children with obstructive sleep apnoea: a
  controlled national study.
\newblock \emph{Thorax}, 68\penalty0 (10):\penalty0 949--954, 2013.

\bibitem[Kim et~al.(2021)Kim, Woo, Jeong, Kim, and Lee]{kim2021sleep}
D.~Kim, Y.~Woo, J.~Jeong, D.-K. Kim, and J.-G. Lee.
\newblock Sleep stage classification for inter-institutional transfer learning.
\newblock In \emph{2021 International Conference on Information and
  Communication Technology Convergence (ICTC)}, pages 1797--1800. IEEE, 2021.

\bibitem[Kushida et~al.(2005)Kushida, Littner, Morgenthaler, Alessi, Bailey,
  Coleman~Jr, Friedman, Hirshkowitz, Kapen, Kramer, and
  {others}]{kushida_practice_2005}
C.~A. Kushida, M.~R. Littner, T.~Morgenthaler, C.~A. Alessi, D.~Bailey,
  J.~Coleman~Jr, L.~Friedman, M.~Hirshkowitz, S.~Kapen, M.~Kramer, and
  {others}.
\newblock Practice parameters for the indications for polysomnography and
  related procedures: an update for 2005.
\newblock \emph{Sleep}, 28\penalty0 (4):\penalty0 499--523, 2005.

\bibitem[Lee et~al.(2022)Lee, Li, DeForte, Splaingard, Huang, Chi, and
  Linwood]{lee2022large}
H.~Lee, B.~Li, S.~DeForte, M.~L. Splaingard, Y.~Huang, Y.~Chi, and S.~L.
  Linwood.
\newblock A large collection of real-world pediatric sleep studies.
\newblock \emph{Scientific Data}, 9\penalty0 (1):\penalty0 1--12, 2022.

\bibitem[Lumeng and Chervin(2008)]{lumeng_epidemiology_2008}
J.~C. Lumeng and R.~D. Chervin.
\newblock Epidemiology of pediatric obstructive sleep apnea.
\newblock \emph{Proc. Am. Thorac. Soc.}, 5\penalty0 (2):\penalty0 242--252,
  2008.

\bibitem[Owens et~al.(2012)Owens, Kothare, and Sheldon]{owens2012pro}
J.~Owens, S.~Kothare, and S.~Sheldon.
\newblock Pro:“not just little adults”: Aasm should require pediatric
  accreditation for integrated sleep medicine programs serving both children
  (0-16 years) and adults.
\newblock \emph{Journal of Clinical Sleep Medicine}, 8\penalty0 (5):\penalty0
  473--476, 2012.

\bibitem[Park and Kim(2022)]{park2022vision}
N.~Park and S.~Kim.
\newblock How do vision transformers work?
\newblock \emph{arXiv preprint arXiv:2202.06709}, 2022.

\bibitem[Paullada et~al.(2021)Paullada, Raji, Bender, Denton, and
  Hanna]{paullada2021data}
A.~Paullada, I.~D. Raji, E.~M. Bender, E.~Denton, and A.~Hanna.
\newblock Data and its (dis) contents: A survey of dataset development and use
  in machine learning research.
\newblock \emph{Patterns}, 2\penalty0 (11):\penalty0 100336, 2021.

\bibitem[Pedregosa et~al.(2011)Pedregosa, Varoquaux, Gramfort, Michel, Thirion,
  Grisel, Blondel, Prettenhofer, Weiss, Dubourg, Vanderplas, Passos,
  Cournapeau, Brucher, Perrot, and Duchesnay]{sklearn}
F.~Pedregosa, G.~Varoquaux, A.~Gramfort, V.~Michel, B.~Thirion, O.~Grisel,
  M.~Blondel, P.~Prettenhofer, R.~Weiss, V.~Dubourg, J.~Vanderplas, A.~Passos,
  D.~Cournapeau, M.~Brucher, M.~Perrot, and E.~Duchesnay.
\newblock Scikit-learn: Machine learning in {P}ython.
\newblock \emph{J. Mach. Learn. Res.}, 12:\penalty0 2825--2830, 2011.

\bibitem[Phan and Mikkelsen(2021)]{phan2021automatic}
H.~Phan and K.~Mikkelsen.
\newblock Automatic sleep staging of eeg signals: Recent development,
  challenges, and future directions.
\newblock \emph{arXiv preprint arXiv:2111.08446}, 2021.

\bibitem[Phan et~al.(2022)Phan, Mikkelsen, Chen, Koch, Mertins, and
  De~Vos]{phan2022sleeptransformer}
H.~Phan, K.~B. Mikkelsen, O.~Chen, P.~Koch, A.~Mertins, and M.~De~Vos.
\newblock Sleeptransformer: Automatic sleep staging with interpretability and
  uncertainty quantification.
\newblock \emph{IEEE Transactions on Biomedical Engineering}, 2022.

\bibitem[Sambasivan et~al.(2021)Sambasivan, Kapania, Highfill, Akrong,
  Paritosh, and Aroyo]{49953}
N.~Sambasivan, S.~Kapania, H.~Highfill, D.~Akrong, P.~K. Paritosh, and L.~M.
  Aroyo.
\newblock "everyone wants to do the model work, not the data work": Data
  cascades in high-stakes ai.
\newblock 2021.

\bibitem[Schreck and Richdale(2011)]{schreck2011knowledge}
K.~A. Schreck and A.~L. Richdale.
\newblock Knowledge of childhood sleep: a possible variable in under or
  misdiagnosis of childhood sleep problems.
\newblock \emph{Journal of sleep research}, 20\penalty0 (4):\penalty0 589--597,
  2011.

\bibitem[Sun et~al.(2017)Sun, Shrivastava, Singh, and Gupta]{sun2017revisiting}
C.~Sun, A.~Shrivastava, S.~Singh, and A.~Gupta.
\newblock Revisiting unreasonable effectiveness of data in deep learning era.
\newblock In \emph{Proceedings of the IEEE international conference on computer
  vision}, pages 843--852, 2017.

\bibitem[Van~der Maaten and Hinton(2008)]{van2008visualizing}
L.~Van~der Maaten and G.~Hinton.
\newblock Visualizing data using t-sne.
\newblock \emph{Journal of machine learning research}, 9\penalty0 (11), 2008.

\bibitem[Vanschoren and Yeung()]{vanschoren2021announcing}
J.~Vanschoren and S.~Yeung.
\newblock Announcing the neurips 2021 datasets and benchmarks track. 2021.
\newblock \emph{URL https://neuripsconf. medium.
  com/announcing-theneurips-2021-datasets-and-benchmarks-track-644e27c1e66c}.

\bibitem[Vaswani et~al.(2017)Vaswani, Shazeer, Parmar, Uszkoreit, Jones, Gomez,
  Kaiser, and Polosukhin]{vaswani2017attention}
A.~Vaswani, N.~Shazeer, N.~Parmar, J.~Uszkoreit, L.~Jones, A.~N. Gomez,
  {\L}.~Kaiser, and I.~Polosukhin.
\newblock Attention is all you need.
\newblock \emph{Advances in neural information processing systems}, 30, 2017.

\bibitem[Watson and Fernandez(2021)]{watson2021artificial}
N.~F. Watson and C.~R. Fernandez.
\newblock Artificial intelligence and sleep: Advancing sleep medicine.
\newblock \emph{Sleep medicine reviews}, 59:\penalty0 101512, 2021.

\bibitem[Yang et~al.(2021)Yang, Wang, Chen, Huang, Ono, Altaf-Ul-Amin, and
  Kanaya]{yang2021exploring}
Z.~Yang, D.~Wang, Z.~Chen, M.~Huang, N.~Ono, M.~Altaf-Ul-Amin, and S.~Kanaya.
\newblock Exploring feasibility of truth-involved automatic sleep staging
  combined with transformer.
\newblock In \emph{2021 IEEE International Conference on Bioinformatics and
  Biomedicine (BIBM)}, pages 2920--2923. IEEE, 2021.

\bibitem[Zhao et~al.(2018)Zhao, Han, Zhang, Wang, Wang, Xu, Tai, Peng, Guo,
  Liu, et~al.]{zhao2018association}
J.~Zhao, S.~Han, J.~Zhang, G.~Wang, H.~Wang, Z.~Xu, J.~Tai, X.~Peng, Y.~Guo,
  H.~Liu, et~al.
\newblock Association between mild or moderate obstructive sleep apnea-hypopnea
  syndrome and cognitive dysfunction in children.
\newblock \emph{Sleep medicine}, 50:\penalty0 132--136, 2018.

\end{thebibliography}

\clearpage
\appendix

\section{Experimental Procedures} \label{sec:methods}
\subsection{Data Description}
\label{subsec_appendix:data}

\begin{figure}[thp]
    \centering
    \includegraphics[width=0.6\linewidth]{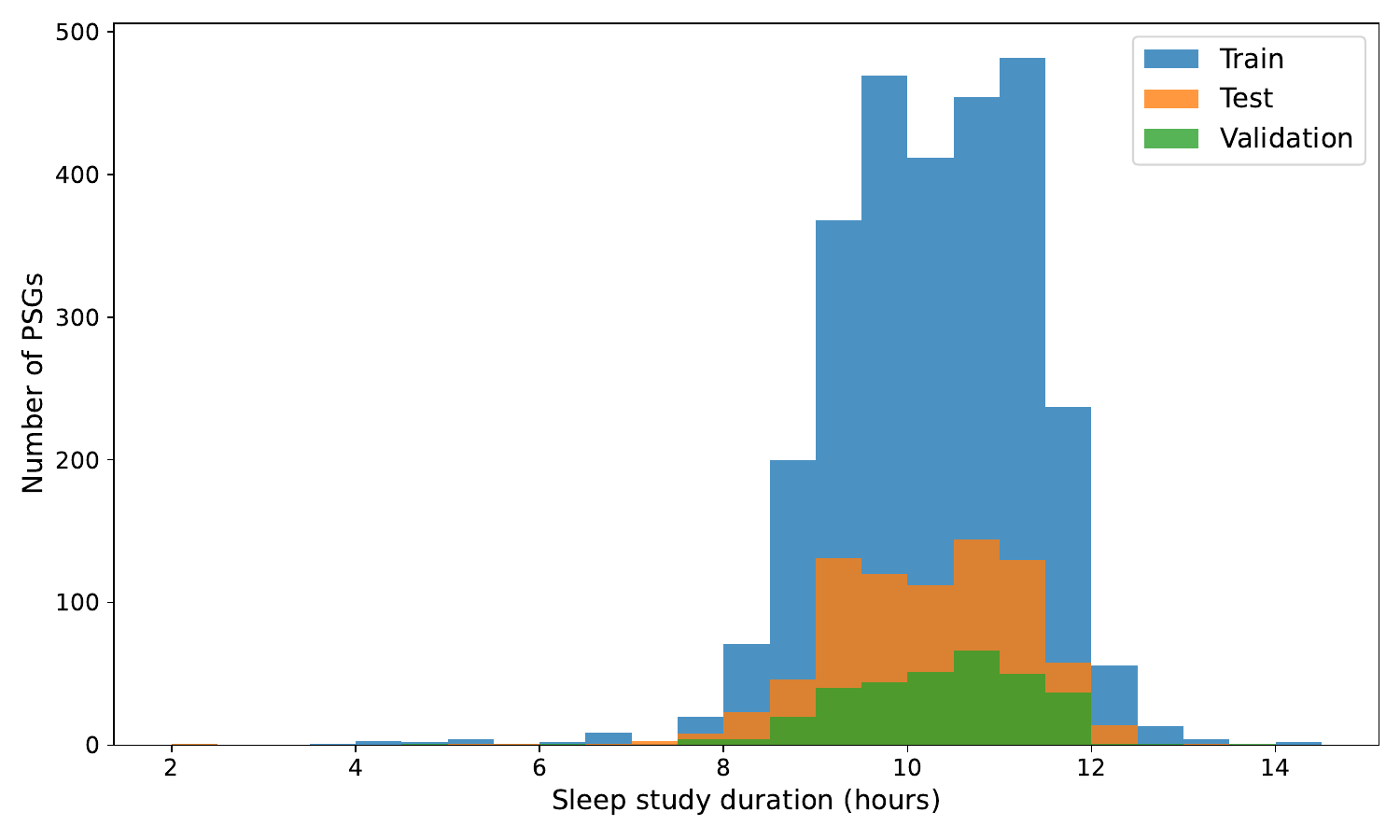}
    \caption{Distributions of sleep study duration in train, test, and validation sets. For all three sets of EDFs, the means were between 10.21 and 10.31 hours, and standard deviations were between 1.05 and 1.09 hours.}
    \label{fig:edf_lengths}
\end{figure}

The NCH Sleep DataBank holds $3,984$ pediatric PSG from 3,673 unique patients that were collected between $2017$ and $2019$ at NCH, Cleveland, Ohio, USA. In this paper, we used $3,928$ PSG from $3,631$ unique patients that had seven EEG channels of interest (F4-M1, O2-M1, C4-M1, O1-M2, F3-M2, C3-M2, CZ-O1), which is about $98.5$\% of the dataset. Demographic information is summarized in Table \ref{tbl:patient_demographics}, and the distributions of sleep study length are visualized in Figure \ref{fig:edf_lengths}. The PSGs were conducted in standard care at NCH, and all sleep stages were manually scored by a technician and verified by a physician board certified in sleep medicine. Since the EEG signals in this dataset have varying sampling frequency, they were resampled to $128$Hz before training the model. Please see \citet{lee2022large} for a much more detailed description of the dataset including the de-identification and validation process. Furthermore, in Table~\ref{tab:dataset_size}, we summarize the size of train, validation and test sets splits.

\begin{table}[htp]
    \caption{Demographic characteristics of the 3,928 PSGs from NCH Sleep DataBank that were used to train, validate, and test our sleep scoring model. $N$ refers to counts; Age is in years; Others and Unknown races include Unknown, Refuse to answer, Native Hawaiian or Other Pacific Islander, and American Indian or Alaska Native, which are aggregated for patient privacy. Note that patients who have gone through multiple sleep studies over the years could have been counted multiple times in different age groups.}
    \label{tbl:patient_demographics}
    \footnotesize
    \centering
    \begin{tabular}{ @{} l rrr @{}} 
    \toprule
    & \multicolumn{3}{r}{\textbf{PSGs, $N$ (Unique Patients, $N$)}} \\ \addlinespace[1mm]
      & Train & Validation  & Test \\
    \midrule \addlinespace[1mm]
     & 2812 (2613) &  321 (291) & 795 (727)  \\ \addlinespace[1mm]
     \textbf{Age} & & &\\  \addlinespace[1mm]
    ~0-1 & 157 (132) & 26 (21)  &  59 (43) \\\addlinespace[1mm]
    ~1-2 & 140 (134) & 15 (14) & 37 (36) \\\addlinespace[1mm]
    ~2-3 & 211 (206) & 31 (30) & 53 (53) \\\addlinespace[1mm]
    ~3-4 & 189 (187) & 31 (31) & 57 (55) \\\addlinespace[1mm]
    ~4-5 & 197 (193) & 16 (15) & 44 (43) \\\addlinespace[1mm]
    ~5-6 & 178 (177) & 16 (16) & 43 (42) \\\addlinespace[1mm]
    ~6-7 & 178 (176) & 16 (16) & 48 (46) \\\addlinespace[1mm]
    ~7-8 & 165 (164) & 17 (17) & 54 (51) \\\addlinespace[1mm]
    ~8-9 & 157 (154) & 14 (14) & 43 (42) \\\addlinespace[1mm]
    ~9-10 & 136 (134) & 18 (18) & 38 (35) \\\addlinespace[1mm]
    ~10-11 & 142 (138) & 7 (7) & 40 (39) \\\addlinespace[1mm]
    ~11-12 & 131 (128) & 8 (7) & 43 (43) \\\addlinespace[1mm]
    ~12-13 & 136 (130) & 9 (9) & 34 (34) \\\addlinespace[1mm]
    ~13-14 & 111 (110) & 17 (17) & 35 (34) \\\addlinespace[1mm]
    ~14-15 & 101 (96) & 15 (14) & 28 (26) \\\addlinespace[1mm]
    ~15-16 & 132 (123) & 13 (12)  & 23 (22) \\\addlinespace[1mm]
    ~16-17 & 118 (113) & 13 (12) & 32 (31) \\\addlinespace[1mm]
    ~17-18 & 93 (89) & 12 (12) & 29 (28) \\\addlinespace[1mm]
    ~18-100 & 140 (131) & 27 (25)  & 55 (46) \\ \addlinespace[2mm]
    \textbf{Race} & & &  \\\addlinespace[1mm]
    ~White & 1855 (1735) & 211 (190) &  531 (481)\\  \addlinespace[1mm]
    ~Black & 581 (536) & 55 (51) & 154 (144) \\  \addlinespace[1mm]
    ~Multiple race & 198 (30) & 33 (30) & 57 (54) \\  \addlinespace[1mm]
    ~Asian &  71 (60) & 11 (9) &  29 (24)\\  \addlinespace[1mm]
    ~Others and Unknown& 107 (97) & 11 (11) & 24 (24) \\  \addlinespace[2mm]
    \textbf{Sex} & & &  \\\addlinespace[1mm]
    ~Male & 1600 (1471) & 185 (166)& 450 (408) \\\addlinespace[1mm]
    ~Female or Unknown & 1212 (1142) & 136 (125) & 345 (319) \\
    \bottomrule
    \end{tabular}
\end{table}

\begin{table}[!htbp]
    \caption{Number of samples in train, validation, and test sets. One sample is a 30-second sleep epoch.}
    \centering
    \footnotesize
    \begin{tabular}{l rrr }
    \toprule
    \textbf{Sleep Stage}& \textbf{Train} & \textbf{Validation} & \textbf{Test}\\  \midrule 
    All    &2,611,845 &301,116 & 731,344\\ \addlinespace[3mm]
     Wake    &469,473 & 56,327 & 135,845 \\ \addlinespace[1mm]
     N1 & 92,615 &9,768 &25,219 \\ \addlinespace[1mm]
     N2 & 990,299 &112,188  & 273,191 \\ \addlinespace[1mm]
     N3 & 623,164 & 71,728 & 176,308 \\ \addlinespace[1mm]
     REM & 436,294  &51,105 & 120,781 \\ \bottomrule
    \end{tabular}
    \label{tab:dataset_size}
\end{table}

\subsection{Self-Attention in Transformers} \label{subsec_appendix:transformer}
\begin{figure}
    \centering
    \includegraphics[width=\textwidth]{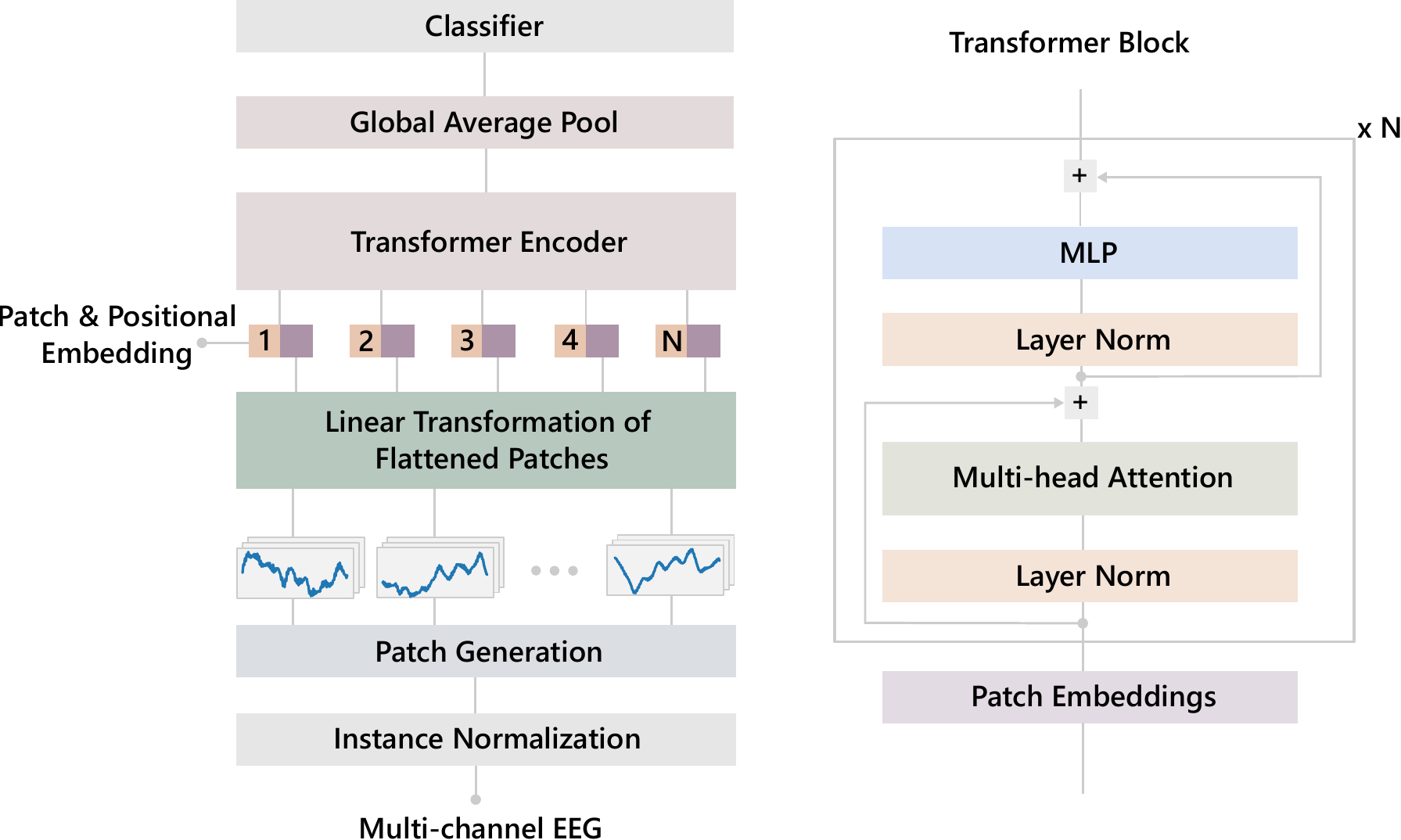}
    \caption{Illustration of our patch-based transformer neural network architecture designed for pediatric sleep scoring from multi-channel EEG signals. MLP stands for multi-layer perceptron.}
    \label{fig:transformer_arch}
\end{figure}

We briefly describe the self-attention mechanism~\citet{vaswani2017attention}, which is a central building block of the transformer architecture. Self-attention computes a weighted average of tokens (or their representation's) with similarity score being equivalent to weights calculated from pairs of tokens. Given an input sequence with multiple channels $X \in \mathbb{R}^{T \times C}$ of length $T$ and channels $C$, it is first reshaped into $n$ patches (or tokens) of fixed size, i.e. $X_p \in \mathbb{R}^{n \times (P \cdot C)}$. Once $X_p$ is projected to $X_t \in \mathbb{R}^{n \times d}$ along with the positional information, it is ready to be inputted into the self-attention module in transformers. The normalized importance matrix is computed using three matrices $W_Q \in \mathbb{R}^{d \times d_q}$, $W_K \in \mathbb{R}^{d \times d_k}$, and $W_V \in \mathbb{R}^{d \times d_v}$, which extract query $Q = X_tW_Q$, key $K = X_tW_K$, and value $V = X_tW_V$. The self-attention is then formulated as:

\begin{equation}
    \mathcal{F}(Q, K, V) = \texttt{Softmax} \big(  \frac{QK^{\top}}{\sqrt{d_q}} \big) V,
\end{equation}

where the softmax operation is applied row-wise, and thus each element in the output matrix depends on all other elements in the same row. Building on top of this, the multi-head self-attention layer comprises $H$ independent self-attention layers. Specifically, each head produce a set of query, key and value matrices and compute attention output as: $h_i = \mathcal{F}(Q_i, K_i, V_i)$ for $i=1, \ldots, H$. Lastly, the fused output is generated by concatenation and linear transformation with learnable weights $W_{O}$:
\begin{equation}
    \mathcal{M}(Q, K, V) = \texttt{Concat}(h_1, h_2, \ldots, h_H) W_{O}.
\end{equation} For a detailed treatment of how multi-head self-attention and transformers work, we refer the reader to~\citet{park2022vision}. In our model, the parameters $T=3840, C=7, n = 30, P=128,  d=d_q=d_k = d_v = 64$, and $H=4$. $T$ is the signal length or temporal size of the instance, $C$ represents the number of channels, $P$ is the patch size, $d$ is the key (including query and value) dimension, and $H$ denotes the number of heads in the multi-head attention layer. The high-level overview of the model architectures is illustrated in Figure~\ref{fig:transformer_arch}.

\subsection{Loss Function}
We use weighted cross-entropy loss function to train our model as NCH data is slightly imbalanced towards N1 class, i.e., there are fewer samples belonging to N1 sleep stage as compared to rest of the classes. Formally, the objective function we optimize is:

\begin{equation}
    \mathcal{L}(\theta) = \frac{1}{M} \sum_{m=1}^{M} \big[ w_m \times \mathbb{H}(y_m, f_{\theta}(y_m | 
    X_m)   \big]
\end{equation}

where $M$ denotes the number of training samples, $X_m$ is the $m$-th EEG instance in the train set, $y_m$ is the $m$-th label in the train set, $f_\theta$ is the neural network function with learnable parameters $\theta$, and $w_m$ is an instance weight representing the importance that should be given to a particular example. In the case of an imbalanced dataset, the $w_m$ is higher for instances from the minority class while lower or one for the rest. For the N1 class, we found the value of $5$ to be optimal, while for the rest of the classes, we used a value of $0.9$ as a weighting factor in the loss function.  $\mathbb{H}$ is the standard cross-entropy loss. The loss function $\mathcal{L}$ is then optimized with respect to the neural network parameters $\theta$ during model training.

\subsection{Model Training and Evaluation}
We use an Adam optimizer with a default learning rate of $0.001$ and batch size of $1,024$ to perform model training on a single NVIDIA T$4$ GPU for $25$ thousand iterations, iterating over more than $2.5$ million multi-channel EEG examples. Our transformer-based model has $775,237$ learnable parameters. We save the model checkpoint at every epoch based on validation set performance to avoid overfitting, and report model performance on the test set. We also experimented with training longer and with an Adam optimizer with weight decay, but we did not notice any improvement in generalization. Finally, we evaluate model performance with four metrics: accuracy, precision, recall, F1-score (macro and weighted averaged variants), and confusion matrix as implemented in the scikit-learn package \citep{sklearn}. Specifically, the F1-score is the harmonic mean of $\texttt{precision}=\frac{TP}{(TP + FP)}$ and $\texttt{recall}=\frac{TP}{(TP+FN)}$, where $TP$ is True Positive, $FP$ is False Positive, and $FN$ is False Negative. In a multi-class classification setting, the macro average is computed as an unweighted mean of per-class F1-scores. In contrast, the weighted average takes each class's support (i.e., number of samples belonging to a particular class) into consideration.  

\subsection{t-SNE Visualization}
\label{subsec_appendix:tsne}

\begin{figure}[!htbp]
\centering
\includegraphics[width=0.45\textwidth]{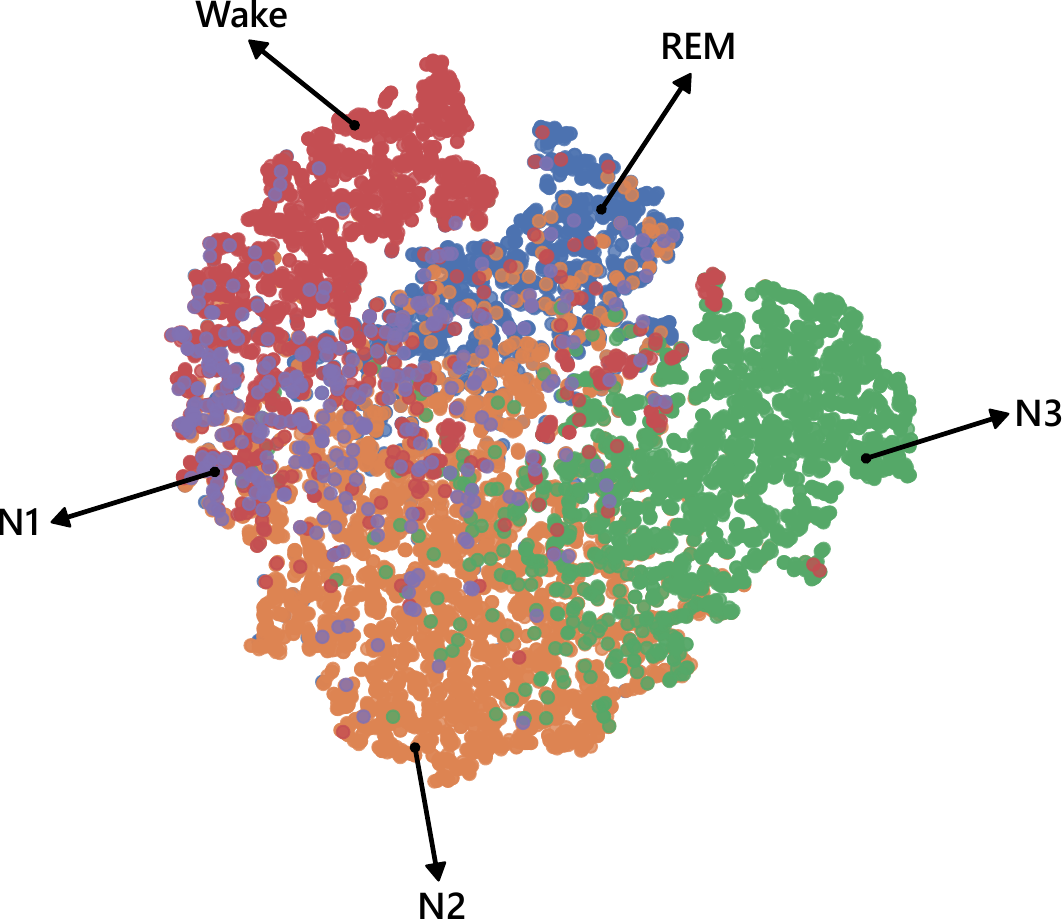}
\caption{t-SNE embedding of features learned with the transformer network.}
\label{fig:tsne_features}
\end{figure}
For a random subset of test set instances, we project $128$-dimensional representations from the model's penultimate layer to $2$ dimensions using t-SNE. That is, each point in Figure \ref{fig:tsne_features} represents one $30$-second sleep epoch on that projected 2 dimensional space. The colors are added during post-hoc analysis for better interpretability, as t-SNE does not utilize class labels. 
\end{document}